\documentclass[12pt,preprint]{aastex}

\usepackage{emulateapj5}
\usepackage{onecolfloat}
\usepackage{times}

\usepackage{aas-macros}
\usepackage{natbib}
\usepackage{graphicx}
\usepackage{dcolumn}
\usepackage{tikz}
\usepackage{pgf}
\usepackage{bm}
\usepackage{amsmath}
\usepackage{amssymb}
\usepackage{hyperref}
\usepackage{float}
\usepackage{booktabs}
\usepackage{tabularx}
\usepackage{gensymb}

\newcommand{\vect}[1]{\bm{#1}}

\shorttitle{Sagittarius stream and the potential of the MW}
\shortauthors{Vera-Ciro \& Helmi}

\begin{document}
\twocolumn[

\title{Constraints on the shape of the Milky Way dark matter halo from the
  Sagittarius stream}

\author{Carlos Vera-Ciro\altaffilmark{1,2} \& Amina
  Helmi\altaffilmark{2}}

\begin{abstract}
  
  We propose a new model for the dark matter halo of the Milky Way that fits
  the properties of the stellar stream associated with the Sagittarius dwarf
  galaxy. Our dark halo is oblate with $q_z = 0.9$ for $r \lesssim 10$ kpc,
  and can be made to follow the \citeauthor{Law2010} model at larger radii.
  However, we find that the dynamical perturbations induced by the Large
  Magellanic Cloud on the orbit of Sgr cannot be neglected when modeling its
  streams.  When taken into account, this leads us to constrain the Galaxy's
  outer halo shape to have minor-to-major axis ratio $(c/a)_\Phi = 0.8$ and
  intermediate-to-major axis ratio $(b/a)_\Phi = 0.9$, in good agreement with
  cosmological expectations.
\end{abstract}

\keywords{galaxies: dwarf - galaxies: interactions - Local Group}]

\altaffiltext{2}{Department of Astronomy, University of Wisconsin, 2535
  Sterling Hall, 475 N. Charter Street, Madison, WI 53076, USA.  {e-mail:
    ciro@astro.wisc.edu}}

\altaffiltext{2}{Kapteyn Astronomical Institute, University of Groningen,
  P.O.Box 800, 9700 AV Groningen, The Netherlands.}

\section{Introduction}
\label{sec:intro}

The stellar stream associated with the Sagittarius (Sgr) dwarf galaxy has been
extensively used to probe the mass distribution of the Milky Way (MW),
particularly its dark halo. Despite many attempts, there is currently no fully
satisfactory model of its shape based on the dynamics of the stream.  Extreme
oblate configurations have been ruled out \citep{Ibata2001}, while the tilt of
the orbital plane has been shown to require a mildly oblate halo by
\cite{Johnston2005}. On the other hand, the line-of-sight velocities call for
a prolate halo \citep{Helmi2004}. This conundrum led \cite[][hereafter
LM10]{Law2010} to propose a triaxial dark halo for the MW, with axis ratios
$(c/a)_\Phi = 0.72 $ and $(b/a)_\Phi = 0.99$ \citep[see also][]{Deg2012}. This
model fits well all positional and kinematic information available.

Although the halos assembled in $\Lambda$CDM are triaxial \citep{Jing2002,
  Allgood2006, Schneider2012}, the configuration proposed by LM10 is rare: the
halo is close to oblate, with a much smaller $c/a$ than predicted in
cosmological simulations for MW mass halos, which have $\langle
c/a\rangle_\Phi=0.9pm 0.1$ over the relevant distance range \citep[i.e., that
probed by the Sgr stream][]{Hayashi2007}, and difficult to understand from a
physical point of view (its minor axis points almost toward the Sun, while the
intermediate axis is perpendicular to the Galactic disk).  Furthermore, the
presence of the disk is expected to lead to a change in the inner halo shape
toward a more oblate configuration \citep{Bryan2012}. Finally, the disk's
stability is not naturally ensured in the LM10 potential, as there are no tube
orbits around the intermediate axis \citep{Debattista2013}.


In this Letter we take a fresh look at determining the shape of the MW halo
from the Sgr streams' dynamics.  We consider the possibility that the shape of
the halo varies with distance from the Galactic center, as expected in
$\Lambda$CDM \citep{Vera2011}.  Evidence suggesting a halo with non-constant
axis ratios has been reported by \cite{Banerjee2011} using the flaring of the
H\textsc{i} layer of the MW disk. We present a new model that takes into
account the effect of a baryonic disk in Section 2. Because of the
cosmological rareness of the LM10 model, in Sectio 3 we explore the
possibility that the dynamics of the Sgr stream may be explained through the
combined effect of the Large Magellanic Could (LMC) and a less axisymmetric,
but more triaxial, outer halo.  In that section we show that these models
provide equally good fits to the dynamics of the young Sgr streams as the LM10
potential, and that older wraps may be used to distinguish amongst them. We
finalize with a brief summary in Section 4.

\section{Inner halo: Accounting for the effect of the Galactic disk on the
  halo shape}
\label{sec:inner-halo}

Next we present the characteristics of our Galactic potential, which includes
a halo whose shape by construction is oblate in the center and triaxial at
large radii. We then show the results of orbital integrations in this
potential aimed at reproducing the properties of the Sgr stream.

\subsection{Description of the Potential}
\label{sec:inner-halo-phi}

We model the Galactic potential with three components: a disk, a spherical
bulge and a dark matter halo. The disk and bulge follow, respectively a
Miyamoto-Nagai distribution \citep[$M_{\rm disk}$ =$ 10^{11}{\rm M}_{\odot}$,
$a$ = 6.5~kpc, $b$ = 0.26~kpc;][]{Miyamotot1975}, and a Hernquist spheroid
\citep[$M_{\rm bulge} = 3.4\times 10^{10}\; {\rm M}_{\odot}$, $c = 0.7$
kpc;][]{Hernquist1990}.
 
Based on the arguments presented in the Introduction, we seek a halo potential
that satisfies the following.
\begin{enumerate}
\item It is axisymmetric in the inner parts. This will guarantee the stability
  of the disk, as well as account for the effects of the baryonic disk on the
  dark halo.
\item It is triaxial in the outskirts, and follows the LM10 model.
\item It has a smooth transition between these two regimes.
\end{enumerate}
We choose to model such a profile using a modification of the algorithm
presented by \cite{Vogelsberger2008}. Consider the spherical potential
\begin{equation}
  \Phi_s(r) = v_{\rm halo}^2\ln(r^2 + d^2).
\end{equation}
The geometrical properties of the potential are encapsulated in the variable
$r=(x^2+y^2+z^2)^{1/2}$. A replacement that satisfies the above requirements
is $r \to \widetilde{r}$, with
\begin{equation}\label{eq:rtilde}
  \widetilde{r} \equiv \frac{r_a + r_T}{r_a + r_A} r_A,
\end{equation}
where $r_A$ and $r_T$ are ellipsoidal radii (as described below). For small
distances $r_A$, $r_T \ll r_a$ then $\widetilde{r}\approx r_A$, and similarly for
large distances $r_A$, $r_T \gg
r_a$ then $\widetilde{r}\approx r_T$.   In particular,
\begin{equation}
 r_A^2 \equiv x^2 + y^2 + \frac{z^2}{q_z^2} = R^2 + \frac{z^2}{q_z^2}, 
\end{equation}
\begin{equation}
\label{eq:r-triaxial}
  r_T^2 \equiv C_1x^2 + C_2y^2 + C_3 xy + \frac{z^2}{q_3^2},
\end{equation}
\noindent and,
\begin{equation}
  C_1 =\frac{a_1^2}{q_1^2} + \frac{a_2^2}{q_2^2}, \quad
  C_2 = \frac{a_1^2}{q_2^2} + \frac{a_2^2}{q_1^2}, \quad
  C_3 =2a_1a_2\left(\frac{1}{q_1^2} - \frac{1}{q_2^2}\right),
\end{equation}
where $a_1 = \cos \phi$ and $a_2 = \sin \phi$, and $\phi = 97^o$. Therefore,
the properties of the mass distribution are encoded in the quantities $r_A$
and $r_T$, with the latter defined as in LM10.  The Sun is assumed to be
located at $x=-R_{\odot}$, and the $z$-axis to point perpendicular to the
disk. The resulting potential,
\begin{equation}\label{eq:potential-halo}
  \Phi_{\rm halo} (x,y,z) = \Phi_s(\widetilde{r}(x, y, z)),
\end{equation}
is axisymmetric at small radii, and triaxial in the outskirts.

\begin{figure}
  \includegraphics[width=0.49\textwidth]{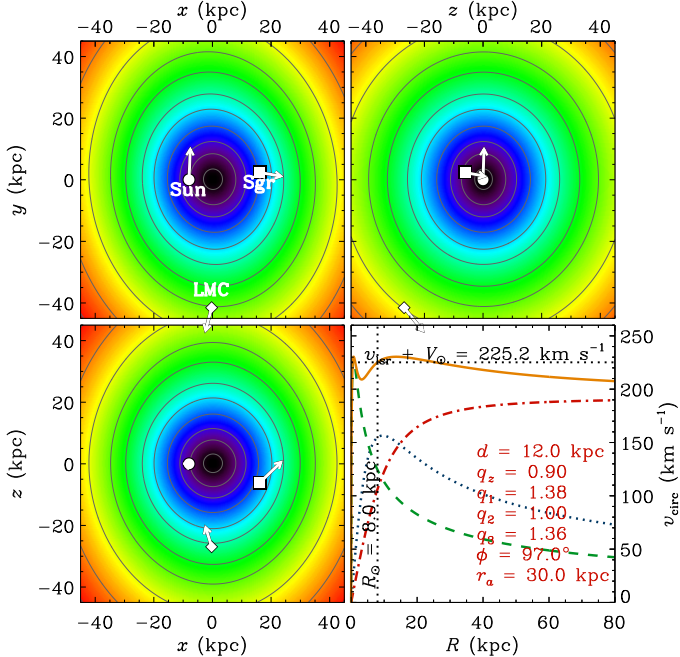}
  \caption{Dark halo potential isocontours on the plane $z = 0$ (top left), $y
    = 0$ (bottom left) and $x = 0$ (top right). For reference, we have
    included the positions and directions of motion for the Sun (circle), Sgr
    (square), and the LMC (diamond). The bottom right panel shows the circular
    velocity profile $v_{\rm circ}$ for the disk (dotted blue), bulge (dashed
    green), and halo (dash dotted red). The halo makes a transition from
    oblate to triaxial at $r_{a} = 30 $ kpc.}
\label{fig:potential-proj}
\end{figure}

Figure~\ref{fig:potential-proj} shows different slices of the resulting
potential. Here we have chosen the flattening of the axisymmetric part to be
$q_z=0.9$ \citep[as in, e.g.,][]{Johnston2005}. The axis ratios for the
triaxial component $(q_1,q_2,q_3)$, and its tilt $\phi$, are taken from the
LM10 model. $v_{\rm halo}$ is set to ensure that $v_{\rm circ} (R_{\odot} =
8{\;\rm kpc}) = 225.2$ km~s$^{-1}$. The transition radius, $r_a = 30$ kpc, is
selected such that the region of dominance of the disk resides inside the
axisymmetric part of the halo potential. However, the effective transition
between the axisymmetric and triaxial regions occurs at a smaller radius,
$\approx 10 $ kpc.

\subsection{Generating the stream}
\label{sec:inner-halo-orbit}

In what follows, we work on the assumption that the orbit of the center of
mass traces the arms of the stream. Although this is not strictly true
\citep{Eyre2009}, it represents a reasonable first approximation
\citep{Law2010}. With this caveat, we proceed to integrate test particles in
the composite potential described above. For each particle, we generate a set
of initial conditions consistent with the present-day six-dimensional (6D)
phase-space coordinates of the Sgr dwarf galaxy. More specifically, we sample
each observable from a Gaussian distribution, with its mean and variance taken
from the literature. The position is assumed to be at $(l, b) = (5\degree.6,
-14\degree.2)$ \citep{Majewski2003}, the heliocentric distance $d = 25 \pm 2 $
kpc \citep{Kunder2009}, the line-of-sight velocity $v_r = 140 \pm 2$ km
s$^{-1}$ \citep{Ibata1997}, and the proper motions $(\mu_l\cos b, \mu_b) =
(-2.4 \pm 0.2, 2.1 \pm 0.2)$ mas yr$^{-1}$ \citep{Dinescu2005}. Orbits are
integrated forward and backward in time for 2 Gyr, to generate the set of
observables associated with the leading and trailing arms, respectively.

For each integrated orbit, we take 10 samples of the form $\{\vect{x}(t_i),
\vect{v}(t_i)\}_{i=1}^{10}$, where the times $t_i$ are randomly selected
between $t=0$ and the maximum time of integration $t_{\rm max}$. $t_{\rm max}$
is the time that it takes the orbit to complete one wrap in the sky, and is
typically $\sim1$ Gyr. The full 6D information contained in each sample is
transformed into the set of observables often used to represent the stream:
position on the sky $(\Lambda_{\odot}, B_{\odot})$ \citep{Majewski2003},
heliocentric distance $d$, line-of-sight velocity in the Galactic standard of
rest $v_{\rm gsr}$, and proper motions $(\mu_b,\mu_l\cos b)$.

\begin{figure}
  \includegraphics[width=0.49\textwidth]{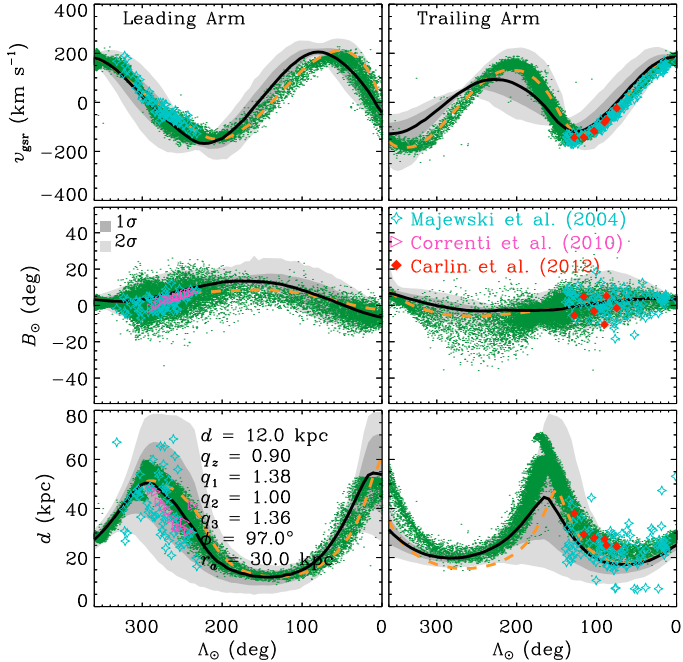}
  \caption{Radial velocity $v_{\rm gsr}$, position in the sky $B_\odot$, and
    heliocentric distance $d$ as function of the angular distance along the
    stream $\Lambda_\odot$ for the leading arm (left) and trailing arm (right)
    for the potential described in Figure~\ref{fig:potential-proj}. The solid
    black line is the median orbit and the shaded regions represent 1$\sigma$
    and 2$\sigma$ equivalent dispersion. The green points are from the
    $N$-body simulation by LM10, while their center of mass orbit is the
    orange dashed curve. }
\label{fig:obs-proj}
\end{figure}

In total $5\times10^4$ initial conditions are integrated, producing
$5\times10^5$ points in the space of observables, which are assigned to a grid
using the Cloud in Cell algorithm
\citep{Hockney1988}. Figure~\ref{fig:obs-proj} shows the projected density for
different observables as a function of $\Lambda_{\odot}$:
$P(o,\Lambda_\odot)$, with $o=\{v_{\rm gsr}$, $B_{\odot}, d, \mu_b, \mu_l\cos
b\}$. In each panel, we marginalize the density over the observed quantity $o$
at fixed $\Lambda_{\odot}$, that is $P(o|\Lambda_{\odot})=\int do
P(o,\Lambda_{\odot})$. The solid black line shows the median of
$P(o|\Lambda_{\odot})$, and with gray bands we represent the 1$\sigma$ and
2$\sigma$ equivalent scatter around the median.

For comparison we have included the mean orbit of the LM10 model (orange
dashed line) and their $N$-body run (green dots). We have also added the
measurements of \citep[][cyan stars]{Majewski2004}, \citep[][magenta
triangles]{Correnti2010} and \citep[][red diamonds]{Carlin2012}. As expected
\citep{Binney2008, Eyre2009}, there are some deviations between the mean orbit
and the location of the tidal stream as probed by the $N$-body run, for
example, in the distances $d$ of the trailing arm.

Figure~\ref{fig:obs-proj} shows that the radial velocities $v_{\rm gsr}$,
distances, and the positions in the sky $B_\odot$ are well fit in our new
potential, and as well as in the LM10 model.  In test runs we found that the
dependence of the fits on the parameter $r_a$ is not strong, whenever this is
kept within reasonable values. Of course, a value of $r_a \gg r_{\rm apo}$
(with $r_{\rm apo}$ the apocenter distance of the orbit of the Sgr dwarf) will
lead to potential that is purely oblate in the region probed by the stream,
and therefore will not be able to fit the velocities of the leading arm.

\begin{figure}
\centering
  \includegraphics[width=0.45\textwidth]{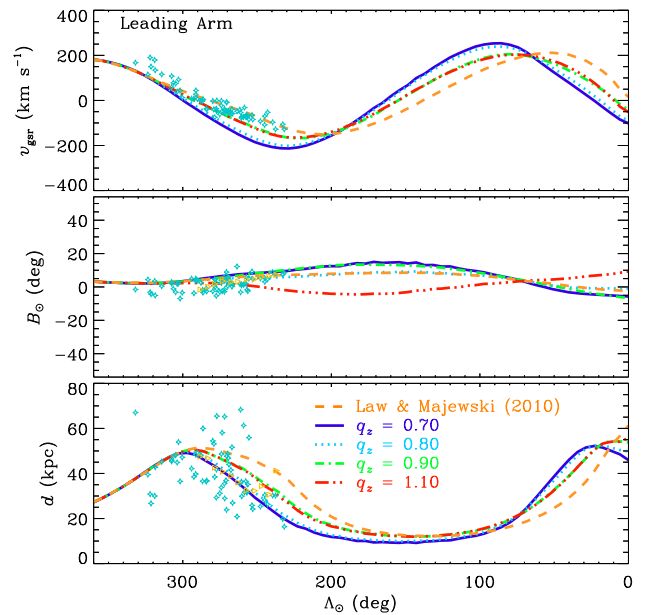}
  \caption{Leading arm line-of-sight velocities (top), position on the sky
    (middle), and heliocentric distances (bottom) for different $q_z$ and
    $r_a=30$ kpc. The potential for $r\gg r_a$ is the same triaxial model as
    in Figure~\ref{fig:obs-proj}. The dashed orange line is the mean orbit of
    LM10.}
\label{fig:obs-proj-all}
\end{figure}

The dependence on the flattening $q_z$, is shown in
Figure~\ref{fig:obs-proj-all} for the leading arm (the trailing arm is rather
insensitive in the region where observations are available). We explore four
different values of $q_z = \{0.7,\; 0.8,\; 0.9,\; 1.1\}$ keeping $r_a = 30$
kpc. In the regions probed by the data, $\Lambda_\odot \gtrsim 200\degree$,
the effect of changing $q_z$ is strong on the velocities, which clearly rule
out $q_z<0.9$.  On the other hand, the positions on the sky disfavor
$q_z>1$. In general, we find that $0.90 < q_z < 0.95$ yield good fits to the
observables in the leading arm. Therefore, Figure~\ref{fig:obs-proj-all} shows
that the inner halo shape has an effect on the Sgr stream, even though the
orbit mainly probes the triaxial regime of the potential.

\section{Outer halo: The effect of LMC}
\label{sec:outer-halo}

It is very intriguing that the direction of the major axis of the LM10
potential approximately lies in the direction toward the LMC. This suggests
that the LM10 potential may perhaps be seen as an effective field: the result
of the combined potentials of the LMC and of a truly triaxial MW halo.

Let us consider the various torques exerted on the (instantaneous) plane of
motion of Sgr. First, note that since $\phi \approx 90\degree$, the principal
axes of the potential of the halo are nearly aligned with the Galactocentric
coordinate system. Consequently, we can simplify
Equation~\eqref{eq:r-triaxial} to
\begin{equation}
  \widetilde{r}^2 \approx x^2 + \frac{y^2}{q_1^2} + 
  \frac{z^2}{q_3^2}.
\end{equation}
The torque induced by the LM10 potential is simply $\vect{\tau} =
-\vect{r}\times \partial \Phi_{\rm halo}/\partial{\vect{r}}$. Of the three
components of this field, the $x$- and $z$- components are controlled by
gradient of the force along the $y$- direction, i.e., that of the major axis of
the LM10 halo. Consider, for instance, the $z$ component,
\begin{eqnarray}\label{eq:torque-halo}
  \tau_z^{\rm halo} &=& 
  -\frac{\partial\Phi_{\rm halo}}{\partial\widetilde{r}} \frac{xy}{\widetilde{r}}
  \left(\frac{1}{q_1^2} - 1 \right) \approx -\frac{v_{\rm circ}^2}{\widetilde{r}^2}xy
  \left(\frac{1}{q_1^2} - 1 \right) \nonumber \\
  &\approx& -\frac{GM_{\rm halo}(\widetilde{r})}{\widetilde{r}^3}xy
  \left(\frac{1}{q_1^2} - 1 \right).
\end{eqnarray}
In our reference system, the present-day position of the LMC is nearly on the
plane $x=0$ (see Figure~\ref{fig:potential-proj}). The force generated at
$\vect{r} = x\vect{i} + y\vect{j} + z\vect{k}$ by a point mass $M_{\rm LMC}$
at the position of the LMC, $\vect{r}_{\rm LMC}$, is
\begin{equation}
  \vect{F}_{\rm LMC} = -GM_{\rm LMC}\frac{\vect{r} - \vect{r}_{\rm LMC}}
  {|\vect{r} - \vect{r}_{\rm LMC}|^3},
\end{equation}
which generates a torque $ \vect{\tau}^{\rm LMC} = \vect{r}\times
\vect{F}_{\rm LMC}$, whose $z$-component is
\begin{equation}\label{eq:torque-lmc}
  \tau_{z}^{\rm LMC} = -\frac{GM_{\rm LMC}}{|\vect{r} - \vect{r}_{\rm LMC}|^3} 
  (yx_{\rm LMC} - xy_{\rm LMC}) \approx \frac{GM_{\rm LMC} xy_{\rm LMC}}
  {|\vect{r} - \vect{r}_{\rm LMC}|^3}.
\end{equation}
Using Equations~(\ref{eq:torque-halo}) and (\ref{eq:torque-lmc}) we can
quantify the relative amplitude of the torques exerted by the triaxial halo
and by the LMC on the orbit of Sgr at its present location:
\begin{equation}
  \frac{\tau_{z}^{\rm LMC}}{\tau_{z}^{\rm halo}} \sim \frac{M_{\rm LMC}}{M_{\rm halo}(\widetilde{r})}
  \frac{\widetilde{r}^3}{r_{\rm sgr/LMC}^3} \frac{y_{\rm LMC}}{y} \frac{1}{1/q_1^2 - 1}.
\end{equation}
The mass of the LM10 halo enclosed at the present distance of Sgr is $M_{\rm
  halo}\sim 10^{11}\;{\rm M}_{\odot} \approx M_{\rm LMC}$
\citep{Besla2010}. At the present day $\widetilde{r}/r_{\rm sgr/LMC}\sim 0.5$,
while $y_{\rm LMC}/y \sim 10$, and taking $q_1 = 1.38$, this implies that the
expression above is of order unity. Additionally, since $q_1 > 1$ the torque
generated by the LMC points in the same direction of that induced by the
triaxial halo ($y_{\rm LMC}<0$, cf. Figure~\ref{fig:potential-proj}). This
means that presently the torque on Sgr generated by the LMC is as important as
the one generated by the triaxial halo.

To confirm this order of magnitude argument we perform new orbital
integrations in a slightly modified halo model, which is still given by
Equation~(\ref{eq:potential-halo}) but now with axis ratios $q_1 = 1.1$, $q_2
= 1.0$, and $q_3 = 1.25$, and where we kept the orientation $\phi =
97\degree$. This corresponds to a minor-to-major axis ratio $(c/a)_\Phi = 0.8$
and intermediate-to-major axis ratio $(b/a)_\Phi = 0.9$, which are consistent
with current predictions of dark matter only simulations of MW type halos
\citep{Hayashi2007}. We also include the potential of the LMC. To this end, we
evolve backward and forward the orbit of the LMC, from its present day
position $(\alpha, \delta) = (5^{\rm h}, 27^{\rm m}.6, -69\degree, 52'.2)$
\citep{Piatek2008}, heliocentric distance $d = 50.1$ kpc \citep{Freedman2001},
proper motions ($\mu_l\cos b$, $\mu_b$) = (1.96, 0.44) mas yr$^{-1}$
\citep{Piatek2008} and line-of-sight velocity $v_r=270$ km s$^{-1}$
\citep{vanderMarel2002}. We then place a Hernquist sphere of mass $M_{{\rm
    LMC}, 0} = 8\times 10^{10}\;{\rm M}_{\odot}$ and scale radius $r_{{\rm
    LMC}, 0} = 2$ kpc along this orbit.

\begin{figure}
  \includegraphics[width=0.49\textwidth]{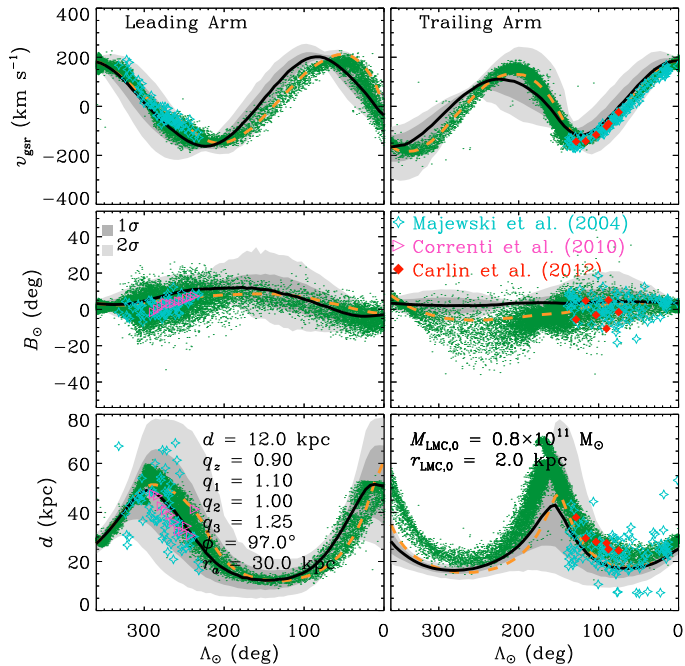}
  \caption{Radial velocity $v_{\rm gsr}$, position in the sky $B_\odot$,
    heliocentric distance $d$ and proper motions $\mu_l\cos b, \mu_b$ as a
    function of the angular distance along the stream $\Lambda_\odot$ for the
    leading arm (left) and the trailing arm (right). The potential used
    includes the LMC as well as that for the halo, which has the form
    described in Equations.~(\ref{eq:rtilde})--(\ref{eq:potential-halo}),
    i.e., it is oblate in the center with $q_z = 0.9$ and $r_a = 30$ kpc, but
    with axis ratios $q_1 = 1.1$, $q_2 = 1.0$, and $q_3 = 1.25$.}
\label{fig:obs-proj-lmc}
\end{figure}

Figure~\ref{fig:obs-proj-lmc} shows a model of the Sgr stream orbital path in
which the LMC and our slightly revised halo are included. As before we show
the LM10 model with the dashed orange line. It is interesting to note that
after including the LMC, the orbital tilt of the leading arm is well fit
despite the change in shape of the MW halo, which is now elongated in the
$z$-direction at large radii. The torque of the orbital plane is also felt by
the trailing arm, resulting on a slight change in the direction of gradient of
$B_{\odot}$ for $\Lambda \lesssim 100\degree$.

This analysis shows that the perturbations of the LMC on the orbit of the Sgr
stream are non-negligible, and implies that previously estimated values of the
axis ratios of the MW dark matter potential from models that have omitted this
perturbation may be biased. For example, $(c/a)_\Phi = 0.8$ for the model
presented in Figure~\ref{fig:obs-proj-lmc}, with the LMC included, while the
LM10 model has $(c/a)_\Phi = 0.72$.

In a more realistic scenario including dynamical friction, the LMC might have
been even more massive than at present day and its role in shaping the orbit
of Sgr even more important. However, some caution is necessary before drawing
strong conclusions about the dynamics of the stream 3-4 Gyr ago. For example,
if the LMC is in its first infall, in which case the closest encounter with
the Sgr dwarf galaxy is currently taking place.  However, during the last
$\sim2$ Gyr its presence could have significantly affected older wraps of the
Sgr stream.

\begin{figure*}
  \includegraphics[width=0.49\textwidth]{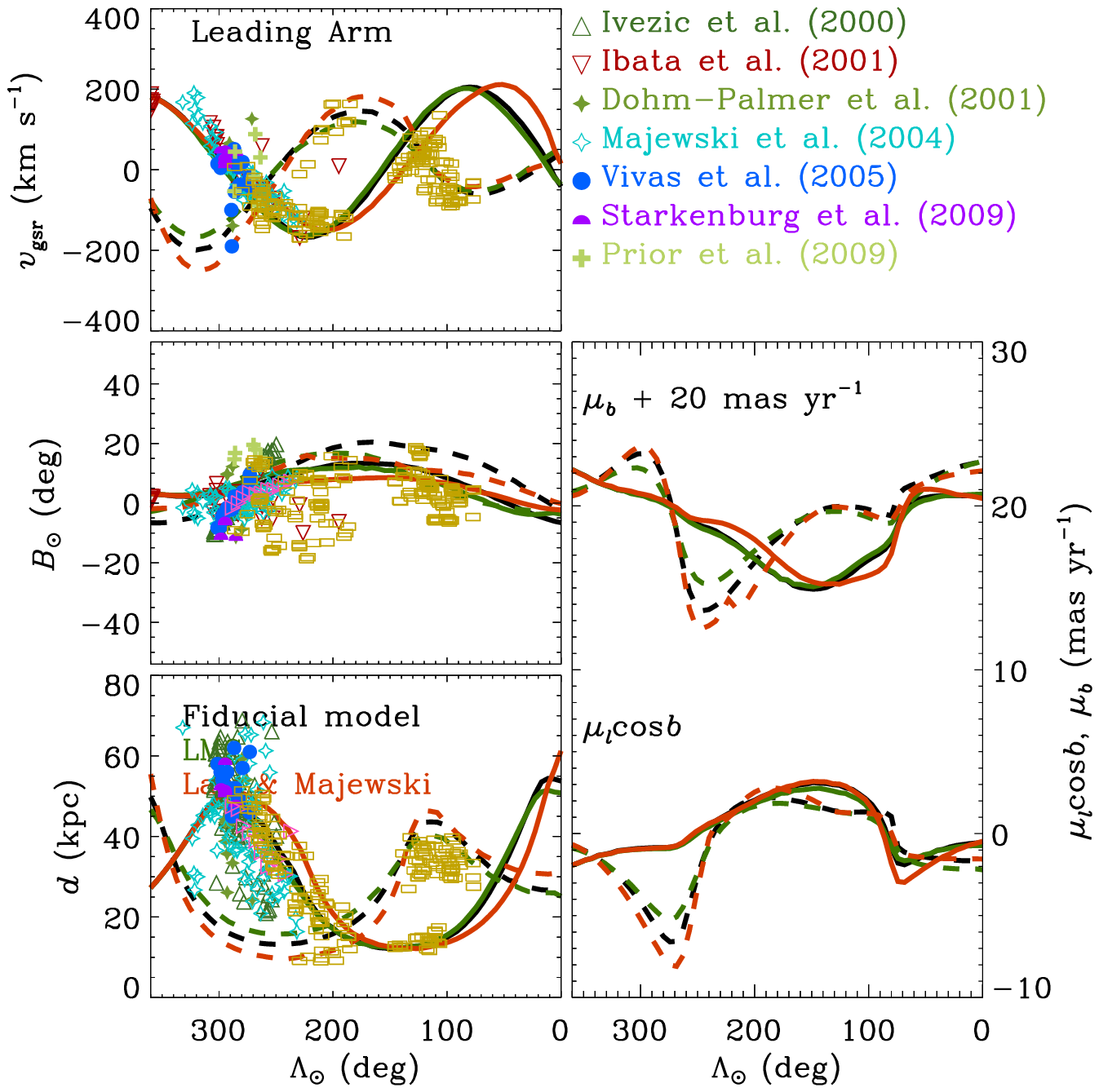}
  \includegraphics[width=0.49\textwidth]{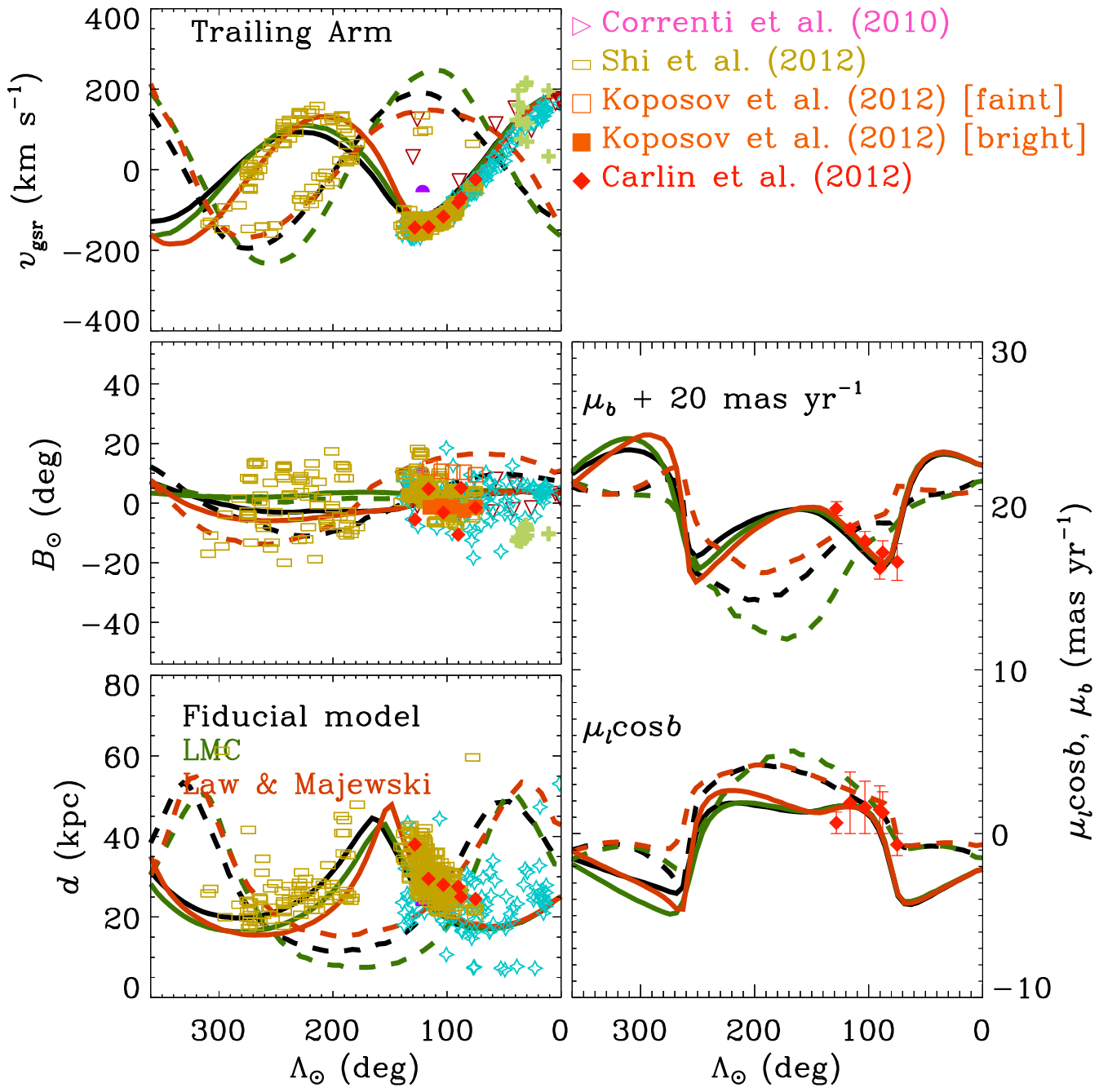}
  \caption{First and second wraps of the leading and trailing streams from Sgr
    for the different models explored: black is our fiducial model from
    Section \ref{sec:inner-halo}, green is the model that includes the LMC and
    red is the LM10 triaxial model.}
\label{fig:second-wrap-models}
\end{figure*}

We explore in Figure~\ref{fig:second-wrap-models} how the differences between
the various models may become apparent for older portions of the stream.  Here
we show the first (solid) and second (dotted) wraps of the leading (left) and
trailing (right) arms, for the three different models discussed so far: black
is our fiducial model from Section \ref{sec:inner-halo}; green is the model
that includes the LMC and red; is the LM10 triaxial model. We have included
observations of different stellar tracers: RR Lyraes \citep{Ivezic2000,
  Vivas2005, Prior2009}, carbon giants \citep{Ibata2001}, red giant branch
stars \citep{Dohm2001, Starkenburg2009, Correnti2010}, M giants
\citep{Majewski2004} and red horizontal branch stars \citep{Shi2012}. It
should be noted that \cite{Shi2012} preselect their sample according to the
LM10 model.

We show also the positions in the sky for the bright (orange filled squares)
and the faint (orange open squares) streams in the Southern Galactic
hemisphere from \cite{Koposov2012}. Whereas the association to the trailing
arm is clear for the brighter portion of the stream, the faint parallel stream
could perhaps be an older wrap from either trailing or leading arm. More
information, especially kinematic, is necessary to disentangle the various
contributions of Sgr in this region, and these might also help constrain
further the shape of the dark halo of the MW. It should be borne in mind that
although the differences between older wraps amongst the various models are
larger than for younger streams, the predictions for their properties are
clearly much more uncertain.

\section{Conclusions}
\label{sec:conclusions}

In this Letter we have presented a new model for the MW dark matter halo that
fits the observations of the Sgr stream, both the radial velocities as well as
the orbital tilt of the leading arm. The dark halo potential is axisymmetric
and flattened toward the disk plane for $r \lesssim 10$ kpc, with $q_z = 0.9$,
and asymptotically approaches the \cite{Law2010} triaxial model at larger
radii.  A gratifying property of this potential is that its inner oblate shape
and orientation account for the presence of the Galactic disk and ensure its
stability.

The triaxial part of this potential, however is not entirely consistent with
expectations from the $\Lambda$CDM model. Its odd (nearly oblate)
configuration can be changed, and brought to a more cosmologically plausible
shape, if the gravitational field generated by the LMC is taken into
account. The integration of orbits in a composite potential including the LMC
and an outer triaxial halo with $q_1 = 1.10$, $q_2 = 1.00$ and $q_3 = 1.25$
(that is, as before, oblate in the inner regions) is also found to reproduce
well the properties of the Sgr streams in the region where these have been
constrained observationally.

The conclusions drawn in this work are based on heuristic searches of the
high-dimensional parameter space that characterizes the gravitational
potential of the MW and that of the LMC. By no means do they represent
best-fit models in a statistical sense. Therefore, the predictions made cannot
be considered exclusive or definitive, but serve to guide where future
observations could focus to distinguish between various
models. Notwithstanding these caveats, we have been able to demonstrate that
the dynamics of the Sgr streams can be understood in a context that is
consistent with expectations from $\Lambda$CDM.


\section*{Acknowledgments}
We are grateful for the financial support from the European Research Council
under ERC-StG grant GALACTICA-240271.

\bibliographystyle{mn2e}

\end{document}